\documentclass[acus,
              ]{jacow}

\makeatletter
\ifboolexpr{bool{xetex}}                 
 {\renewcommand{\Gin@extensions}{.pdf,%
                    .png,.jpg,.bmp,.pict,.tif,.psd,.mac,.sga,.tga,.gif,%
                    .eps,.ps,%
                    }}{}
\makeatother

\ifboolexpr{bool{xetex} or bool{luatex}} 
 {}                                      
 {\usepackage[utf8]{inputenc}}           

\usepackage[USenglish]{babel}

\ifboolexpr{bool{jacowbiblatex}}
 {%
  \addbibresource{jacow-test.bib}
  \addbibresource{biblatex-examples.bib}
 }{}

\newcommand{\q}[2]{\ensuremath{#1\ \mathrm{#2}}} 

\newcommand{\vol}[1]{\textbf{#1}} 
\newcommand{\kick}{\ensuremath{\theta}}
\newcommand{\betaCS}{\ensuremath{\beta_\mathrm{lat}}} 
\newcommand{\Brho}{\ensuremath{(B\rho)}} 

\begin{document}

\title{Electron Lenses and Cooling for the Fermilab Integrable Optics Test
Accelerator\thanks{Fermilab is operated by Fermi Research Alliance, LLC
under Contract No.~DE-AC02-07CH11359 with the United States Department
of Energy. Report number: FERMILAB-CONF-15-446-AD-APC.}}

\author{G.~Stancari\thanks{Email: $\langle$stancari@fnal.gov$\rangle$.},
A.~Burov, V.~Lebedev, S.~Nagaitsev, E.~Prebys, A.~Valishev, Fermilab, Batavia IL, USA}

\maketitle

\begin{abstract}
Recently, the study of integrable Hamiltonian systems has led to
nonlinear accelerator lattices with one or two transverse invariants and
wide stable tune spreads. These lattices may drastically improve the
performance of high-intensity machines, providing Landau damping to
protect the beam from instabilities, while preserving dynamic aperture.
The Integrable Optics Test Accelerator (IOTA) is being built at Fermilab
to study these concepts with 150-MeV pencil electron beams
(single-particle dynamics) and 2.5-MeV protons (dynamics with self
fields). One way to obtain a nonlinear integrable lattice is by using
the fields generated by a magnetically confined electron beam (electron
lens) overlapping with the circulating beam. The required parameters are
similar to the ones of existing devices. In addition, the electron lens
will be used in cooling mode to control the brightness of the proton
beam and to measure transverse profiles through recombination. More
generally, it is of great interest to investigate whether nonlinear
integrable optics allows electron coolers to exceed limitations set by
both coherent or incoherent instabilities excited by space charge.
\end{abstract}

\section{Introduction}

In many areas of particle physics, such as the study of neutrinos and of
rare processes, high-power accelerators and high-brightness beams are
needed. The performance of these accelerators is limited by several
factors, including tolerable losses and beam halo, space-charge effects,
and instabilities. Nonlinear integrable optics, self-consistent or
compensated dynamics with self fields, and beam cooling beyond the
present state of the art are being actively pursued because of their
potential impact.

\begin{figure}[b!]
\includegraphics[width=\columnwidth]{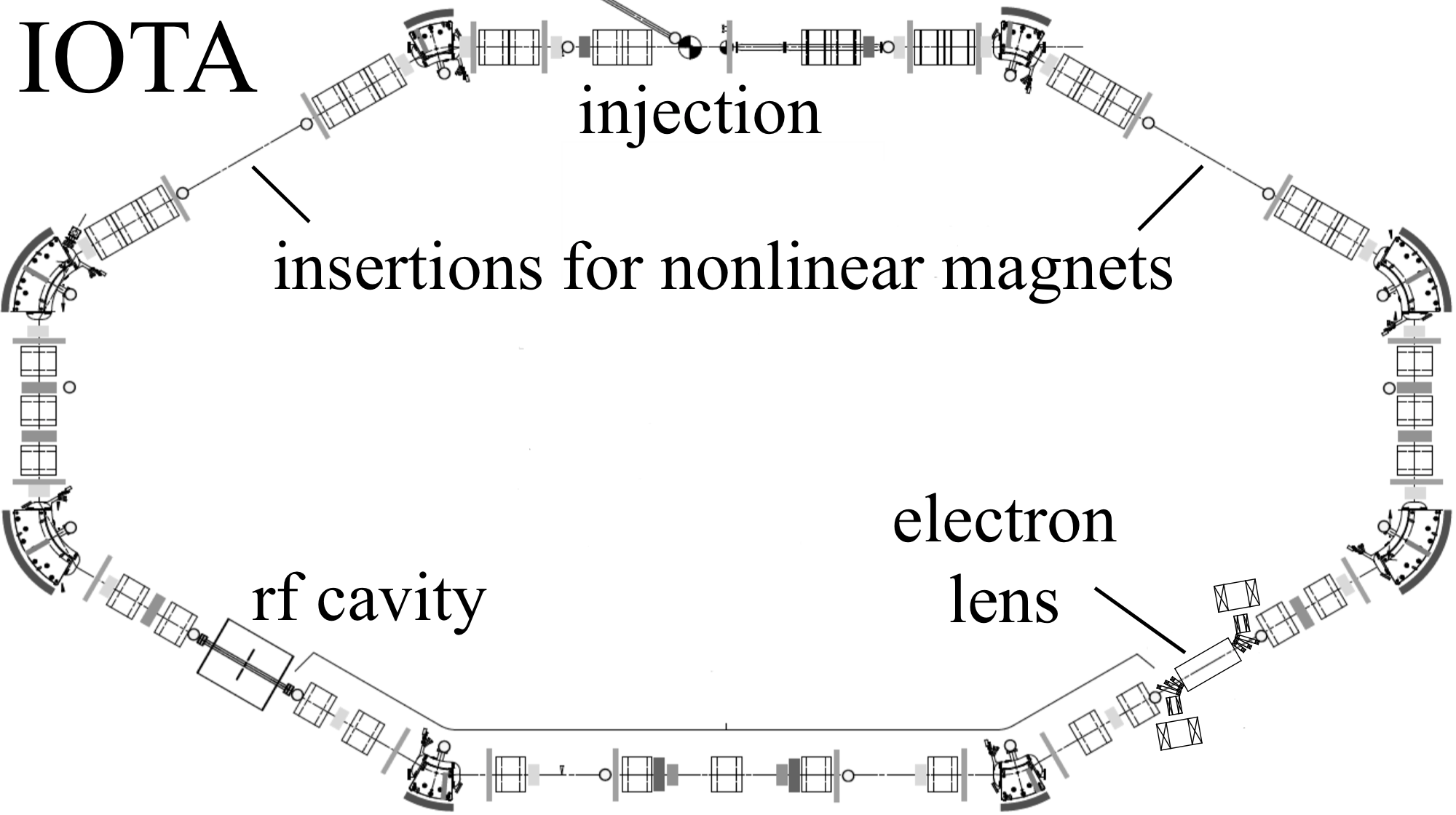}
\caption{Layout of the IOTA ring.}
\label{fig:IOTA}
\end{figure}

In particular, the Integrable Optics Test Accelerator (IOTA,
Fig.~\ref{fig:IOTA}) is a research storage ring with a circumference of
40~m being built at Fermilab~\cite{Nagaitsev:IPAC:2012,
Valishev:IPAC:2012}. Its main purposes are the practical implementation
of nonlinear integrable lattices in a real machine, the study of
space-charge compensation in rings, and a demonstration of optical
stochastic cooling. IOTA is designed to circulate pencil beams of
electrons at 150~MeV for the study of single-particle linear and
nonlinear dynamics. For experiments on dynamics with self fields,
protons at 2.5~MeV (momentum 69~MeV/$c$) will be used.

In accelerator physics, nonlinear integrable optics involves a small
number of special nonlinear focusing elements added to the lattice of a
conventional machine in order to generate large tune spreads while
preserving dynamic aperture~\cite{Danilov:PRSTAB:2010}, thus providing
improved stability to perturbations and mitigation of collective
instabilities through Landau damping.

One way to generate a nonlinear integrable lattice is with specially
segmented multipole magnets~\cite{Danilov:PRSTAB:2010}. There are also
two concepts based on electron lenses~\cite{Stancari:IPAC:2015}:
(a)~axially symmetric thin kicks with a specific amplitude
dependence~\cite{McMillan:1967, McMillan:1971, Danilov:PAC:1997}; and
(b)~axially symmetric kicks in a thick lens at constant amplitude
function~\cite{Nagaitsev:private, Stancari:AAC:2014}. These concepts use
the electromagnetic field generated by the electron beam distribution to
provide the desired nonlinear transverse kicks to the circulating beam.

In IOTA, the electron lens can also be used as an electron cooler for
protons. In this paper, we present a preliminary exploration of the
research opportunities enabled by the cooler option: beam dynamics with
self fields can be studied in a wider brightness range; spontaneous
recombination provides fast proton diagnostics; and, lastly, perhaps the
most interesting question is whether the combination of electron cooling
and nonlinear integrable optics leads to higher brightnesses than
presently achievable.

\section{Nonlinear integrable optics with electron lenses}

Electron lenses are pulsed, magnetically confined, low-energy electron
beams whose electromagnetic fields are used for active manipulation of
circulating beams~\cite{Shiltsev:PRSTAB:2008, Shiltsev:Handbook:2013}.
One of the main features of an electron lens is the possibility to
control the current-density profile of the electron beam (flat,
Gaussian, hollow, etc.) by shaping the cathode and the extraction
electrodes. Electron lenses have a wide range of
applications~\cite{Shiltsev:elens-bbcomp, Shiltsev:PRL:2007,
Zhang:PRSTAB:2008, Stancari:BB:2013, Stancari:PRL:2011,
Stancari:APSDPF:2011, Fischer:IPAC:2013, Fischer:IPAC:2014,
Stancari:CDR:2014, Valishev:TM:2013, Valishev:IPAC:2015:hllhc}. In
particular, they can be used as nonlinear lenses with tunable kicks and
controllable shape as a function of betatron amplitude.

The goal of the nonlinear integrable optics experiments, including the
ones with electron lenses, is to achieve a large tune spread, of the
order of 0.25 or more, while preserving the dynamic aperture and
lifetime of the circulating beam. Experimentally, this will be observed
by recording the lifetime and turn-by-turn position of a low-intensity,
low-emittance 150-MeV circulating electron bunch, injected and kicked to
different betatron amplitudes, for different settings of the nonlinear
elements (magnets or electron lenses).

The cathode-anode voltage~$V$ determines the velocity $v_e = \beta_e c$
of the electrons in the device, which is assumed to have length~$L$ and
to be located in a region of the ring with lattice amplitude
function~\betaCS. When acting on a circulating beam with magnetic
rigidity~\Brho\ and velocity $v_z = \beta_z c$, the linear focusing
strength~$k_e$ for circulating particles with small betatron amplitudes
is proportional to the electron current density on axis~$j_0$:
\begin{equation}
k_e = 2\pi \frac{j_0 L (1 \pm \beta_e \beta_z)}
                {\Brho \beta_e \beta_z c^2}
            \left( \frac{1}{4\pi\epsilon_0} \right) .
\label{eq:ke}
\end{equation}
The `$+$' sign applies when the beams are counter-propagating and the
electric and magnetic forces act in the same direction. For small
strengths and away from the half-integer resonance, these kicks
translate into the tune shift
\begin{equation}
\Delta\nu = \frac{\betaCS j_0 L (1 \pm \beta_e \beta_z)}
                {2 \Brho \beta_e \beta_z c^2}
            \left( \frac{1}{4\pi\epsilon_0} \right) .
\label{eq:dnu}
\end{equation}
for particles circulating near the axis.

There are two concepts of electron lenses for nonlinear integrable
optics.

\subsection{Thin Radial Kick of McMillan Type}

The integrability of axially symmetric thin-lens kicks was studied in
1~dimension by McMillan~\cite{McMillan:1967, McMillan:1971}. It was then
extended to 2~dimensions~\cite{Danilov:PAC:1997} and experimentally tested
with colliding beams~\cite{Shwartz:BB:2013}. Let $j(r)$ be a specific radial
dependence of the current density of the electron-lens beam, with $j_0$ its
value on axis and $a$ its effective radius:
$j(r) = j_0 a^4 / (r^2 + a^2)^2$.
The total current is
$I_e = j_0 \pi a^2$.
The circulating beam experiences nonlinear transverse kicks:
$\kick(r) = k_e a^2 r / (r^2 + a^2)$.
For such a radial dependence of the kick, if the element is thin
($L\ll\betaCS$) and if the betatron phase advance in the rest of the
ring is near an odd multiple of $\pi/2$, there are 2 independent
invariants of motion in the 4-dimensional transverse phase space.
Neglecting longitudinal effects, all particle trajectories are regular
and bounded. The achievable nonlinear tune spread~$\Delta\nu$ (i.e., the
tune difference between small and large amplitude particles) is of the
order of $\betaCS k_e / 4\pi$ (Eq.~\ref{eq:dnu}). A more general
expression applies when taking into account machine coupling and the
electron-lens solenoid. For the thin McMillan lens, it is critical to achieve and preserve the desired current-density profile.

\subsection{Axially Symmetric Kick in Constant Beta Function}

The concept of axially symmetric thick-lens kicks relies on a section of
the ring with constant and equal amplitude functions. This can be
achieved with a solenoid with axial field~$B_z = 2 \Brho / \betaCS$ to
provide focusing for the circulating beam and lattice functions $\betaCS
\equiv \beta_x = \beta_y$. The same solenoid magnetically confines the
low-energy beam in the electron lens. In this case, any axially
symmetric electron-lens current distribution $j(r)$ generates
2~conserved quantities (the Hamiltonian and the longitudinal component
of the angular momentum), as long as the betatron phase advance in the
rest of the ring is an integer multiple of $\pi$. At large electron beam
currents in the electron lens, the focusing of the electron beam itself
dominates over the solenoid focusing and can be chosen to be the source
of the constant amplitude functions. Because the machine operates near
the integer or half integer resonances, the achievable tune spread in
this case is of the order of $L / (2\pi\betaCS)$. This scenario favors
thick lenses and it is insensitive to the current-density distribution
in the electron lens.

Several operating scenarios for the IOTA electron lenses are possible
within the currently available parameter
space~\cite{Stancari:IPAC:2015}. The feasibility and robustness of these
designs against deviations from the ideal cases are being studied with
analytical calculations and numerical tracking simulations.

\begin{table}[bt]
\centering
\caption{Typical Electron-Lens Parameters for IOTA}
\begin{tabular}{lr}
\toprule
Parameter & Value\\
\midrule
Cathode-anode voltage, $V$ & 0.1--10 kV\\
Beam current, $I_e$ & 5~mA -- 5~A \\
Current density on axis, $j_0$ & 0.1--12 A/cm$^2$ \\
Main solenoid length, $L$ & 0.7 m \\
Main solenoid field, $B_z$ & 0.1--0.8 T\\
Gun/collector solenoid fields, $B_g$ & 0.1--0.4 T\\
Max. cathode radius, $(a_g)_\mathrm{max}$ & 15 mm \\
Amplitude function, \betaCS & 0.5--10~m \\
Circulating beam size (rms), $\sigma_e$ & 0.1--0.5 mm ($e^-$)\\
 & 1--5 mm ($p$) \\
\bottomrule
\end{tabular}
\label{tab:par}
\end{table}

Typical electron-lens parameter ranges for IOTA are shown in
Table~\ref{tab:par}.

\section{Electron Cooling in IOTA}

We investigate the benefits of an electron cooler in the ring and the
possible difficulties of running an electron lens in cooling
configuration.

Electron cooling in IOTA would extend the range of available
brightnesses for space-charge experiments with protons. It would also
provide a flow of neutral hydrogen atoms through spontaneous
recombination for beam diagnostics downstream of the electron lens. Of
greater scientific interest is the question of whether nonlinear
integrable optics allows cooled beams to exceed the limitations of
space-charge tune spreads and instabilities. Here we discuss these three
aspects in more detail.

\subsection{Electron Cooling of Protons}

\begin{table}[bt]
\centering
\caption{Proton Parameters for Cooling in IOTA.}
\begin{tabular}{lr}
\toprule
Parameter & Value \\
\midrule
Kinetic energy, $T_p$ & 2.5~MeV \\
Normalized velocity, $\beta_p$ & 0.073 \\
Number of particles, $N_p$ & $5\times 10^{9}$ \\
Beam current, $I_p$ & 0.44~mA \\
Normalized rms emittance, $\epsilon_{pn}$ & \q{0.3 \to 0.03}{\mu m} \\
Rms beam size at cooler, $y_p$ & 4$\to$1.3~mm \\
Momentum spread, $\sigma_p/p$ & $5\times 10^{-4}$ \\
Space-charge tune shift, $\Delta\nu_\mathrm{sc}$ & $-0.028 \to -0.28$ \\
Transv. temperature (avg.), $\langle k T_{p\perp}\rangle$ & 5$\to$0.5~eV \\
Long. temperature, $k T_{p \parallel}$ & 0.6~eV \\
\bottomrule
\end{tabular}
\label{tab:protons}
\end{table}

Proton parameters are shown in Table~\ref{tab:protons}. The parameters
are chosen to balance the dominant heating and cooling mechanisms, while
achieving significant space-charge tune shifts. To match the proton
velocity, the accelerating voltage in the electron lens has to be $V =
\q{1.36}{kV}$. 

At these energies, proton lifetime is dominated by residual-gas
scattering and by intrabeam scattering, due to emittance growth in the
absence of cooling. (Charge neutralization is discussed below.) At the
residual gas pressure of \q{10^{-10}}{mbar}, the lifetime contributions
of emittance growth due to multiple Coulomb scattering and of losses
from single Coulomb scattering are 40~s and 40~min, respectively.

Intrabeam scattering has a stronger effect. Whereas the transverse
emittance growth time is 120~s, the longitudinal growth time can be as
small as 2.5~s, indicating a possible heat transfer from the
longitudinal to the transverse degrees of freedom, which must be
mitigated by keeping the effective longitudinal temperature of the
electrons (which is dominated by the space-charge depression and
therefore by the density $n_e$) low enough. At the same time, one needs
to ensure that the heating term of the magnetized cooling force is
negligible. One can achieve cooling rates of about 20~ms and reduce the
transverse emittance by about a factor~10, with a corresponding increase
in brightness.

\subsection{Diagnostics through Recombination}

IOTA is a research machine and diagnostics is critical to study beam
evolution over the time scales of instability growth. The baseline
solution for profile measurement consists of ionization monitors, with
or without gas injection. In IOTA, with $N_p = 5\times 10^{9}$
circulating protons, for a residual gas pressure of \q{10^{-10}}{mbar},
one can expect 9~ionizations per turn, or a ionization rate of 4.9~MHz.

Spontaneous recombination $p + e^- \to H^0 + h\nu$ has proven to be a
useful diagnostics for optimizing the cooler settings and to determine
the profile of the circulating beam.

Neutral hydrogen is formed in a distribution of excited Rydberg states,
which have to survive Lorentz stripping through the electron lens toroid
and through the next ring dipole to be detected. For IOTA parameters and
magnetic fields, atomic states up to $n = 12$ can survive. The
corresponding recombination coefficient is $\alpha_r = \q{9.6\times
10^{-19}}{m^3/s}$ for $\sqrt{kT_e} = \q{0.1}{eV}$ (and scales as
$1/\sqrt{kT_e}$).

The total recombination rate~$R$ is also proportional to the fraction of
the ring occupied by the cooler, $L/C = (\q{0.7}{m})/(\q{40}{m})$ and to
the electron density, $n_e$:
\begin{equation}
R = N_p \alpha_r n_e (L/C) (1/\gamma^2)
\label{eq:recombination}
\end{equation}
For $N_p = 5\times 10^{9}$ and $n_e = \q{5.8\times 10^{14}}{m^{-3}}$,
one obtains a rate $R = \q{48}{kHz}$, which is small enough not to
significantly affect beam lifetime, but large enough for relatively fast
diagnostics complementary to the ionization profile monitors.

\subsection{Electron Cooling and Nonlinear Integrable Optics}

A new research direction is suggested by these studies: in the cases
where electron cooling is limited by instabilities or by space-charge
tune spread, does nonlinear integrable optics combined with cooling
enable higher brightnesses? It seems feasible to investigate this
question experimentally in IOTA.

The more straightforward scenario includes electron cooling parameters
such as the ones described above. Integrability and tune spreads are
provided separately by the nonlinear magnets. Space-charge tune spreads
of 0.25 or more, and comparable nonlinear tune spreads, are attainable.

An appealing but more challenging solution would be to combine in the
same device, the electron lens, both cooling and nonlinearity (a lens of
the McMillan type, for instance). If successful, such a solution would
have a direct impact on existing electron coolers in machines that are
flexible enough to incorporate the linear part of the nonlinear
integrable optics scheme (the T-insert described in
Ref.~\cite{Danilov:PRSTAB:2010}). Preliminary studies indicate that it
is challenging to incorporate both the constraints of cooling and the
high currents needed to achieve sizable tune spreads, unless one can
suppress the space-charge depression. This option is still under study.

As a general comment, we add that instabilities are often driven by
impedance. In a research machine dedicated to high-brightness beams, it
is useful to be able to vary the electromagnetic response of the beam
environment. For this reason, positive feedback with a transverse damper
system is being proposed to explore the stability of cooled and uncooled
beams with self fields in linear and nonlinear lattices.

\section{Conclusions}

In the Fermilab Integrable Optics Test Accelerator, nonlinear lenses
based on magnetically confined electron beams will be used for
experimental tests of integrable transfer maps.

With circulating protons, electron lenses can also be used as electron
coolers. Cooling times of less than a second can be achieved, allowing
one to access a wider range of equilibrium brightnesses for the planned
experiments of beam dynamics with self fields.

A recombination detector downstream of the electron lens will complement
ionization monitors for measurements of transverse parameters and
instabilities.

An electron cooler in the nonlinear integrable lattice also enables new
research on the nature of brightness limits for high-intensity cooled
beams. Having the electron lens act both as nonlinear element and as
cooler seems challenging. However, one can rely on the IOTA nonlinear
magnets for stable tune spread generation. In addition, the damper
system will enable research on beam stability with controlled
excitations.

\section{Acknowledgments}

The authors would like to acknowledge the contributions of
D.~Broemmelsiek, K.~Carlson, D.~Crawford, W.~Johnson, M.~McGee,
L.~Nobrega, H.~Piekarz, A.~Romanov, J.~Ruan, V.~Shiltsev, J.~Thangaraj,
R.~Thurman-Keup, T.~Zolkin (Fermilab), S.~Antipov (University of Chicago
and Fermilab) in developing theoretical concepts and technical solutions
for the Fermilab Integrable Optics Test Accelerator.

%
%
\iftrue   
	\raggedend
\fi

\iffalse  
	\newpage
	\printbibliography

\else


\fi

\end{document}